\documentclass[reprint]{revtex4-1}

\usepackage{graphicx}

\begin{document}

\title{Tunable open-access microcavities for on-chip cQED}

\author{C.A. Potts}
\affiliation{ECE Department, University of Alberta, T6G 2V4, Edmonton, AB, Canada}
\affiliation{Department of Physics, University of Alberta, T6G 2E1 Edmonton, AB, Canada}
\author{A. Melnyk}
\affiliation{ECE Department, University of Alberta, T6G 2V4, Edmonton, AB, Canada}
\author{H. Ramp}
\affiliation{Department of Physics, University of Alberta, T6G 2E1 Edmonton, AB, Canada}
\author{M.H. Bitarafan}
\affiliation{ECE Department, University of Alberta, T6G 2V4, Edmonton, AB, Canada}
\author{D. Vick}
\affiliation{National Institute for Nanotechnology, T6G 2M9, Edmonton, AB, Canada}
\author{L.J. LeBlanc}
\affiliation{Department of Physics, University of Alberta, T6G 2E1 Edmonton, AB, Canada}
\affiliation{Canadian Institute for Advanced Research, Toronto, M5G 1Z8 Canada}
\author{J.P Davis}
\email[]{jdavis@ualberta.ca}
\affiliation{Department of Physics, University of Alberta, T6G 2E1 Edmonton, AB, Canada}
\author{R.G. DeCorby}
\email[]{rdecorby@ualberta.ca}
\affiliation{ECE Department, University of Alberta, T6G 2V4, Edmonton, AB, Canada}

\date{\today}

\begin{abstract}
We report on the development of on-chip microcavities and show their potential as a platform for cavity quantum electrodynamics experiments. Microcavity arrays were formed by the controlled buckling of SiO$_2$/Ta$_2$O$_5$ Bragg mirrors, and exhibit a reflectance-limited finesse of 3500 and mode volumes as small as 35$\lambda^3$.   We show that the cavity resonance can be thermally tuned into alignment with the D2 transition of $^{87}$Rb, and outline two methods for providing atom access to the cavity.  Owing to their small mode volume and high finesse, these cavities exhibit single-atom cooperativities as high as $C_1 = 65$. A unique feature of the buckled-dome architecture is that the strong-coupling parameter $g_0/\kappa$ is nearly independent of the cavity size. Furthermore, strong coupling should be achievable with only modest improvements in mirror reflectance, suggesting that these monolithic devices could provide a robust and scalable solution to the engineering of light-matter interfaces.
\end{abstract}

\maketitle
The implementation of a distributed quantum network could enable a global quantum communication system,\cite{Kimble2008,VanEnk1997} distributed quantum computation,\cite{Cirac1997} distribution of quantum entanglement,\cite{Duan2001} and may even provide a global time standard by embedding atomic clocks at quantum nodes.\cite{Komar2014}  To realize these objectives, coherent control of individual quantum states, and their interactions, is required.  Since photons - the prototypical distributed qubit - exhibit no interactions with themselves in vacuum, \cite{Chang2014} matter systems must act as intermediaries to perform quantum-state operations, \cite{Reiserer2014} serve as quantum memories,\cite{Lvovsky2009} and interface with quantum processors.\cite{Petrosyan2009}

Numerous systems have been investigated as quantum light-matter interfaces including phonons in optomechanical devices, \cite{Palomaki2013,Bochmann2013} rare-earth-ion doped crystals,\cite{Zhong2015,Saglamyurek2015} nitrogen vacancies in diamond,\cite{Faraon2011,Li2014,Burkard2014} quantum dots,\cite{Press2008,Greve2012} and alkali gases such as rubidium \cite{Wilk2007,Petrosyan2009} and cesium. \cite{Duan2003} Of particular interest are single alkali-atoms trapped within high-finesse optical cavities, in the context of cavity quantum electrodynamics (cQED). \cite{Walther2006,Mabuchi2002}  Within the past decade, single atoms trapped in cavities have been used to store quantum information, \cite{Specht2011,Nolleke2013,Kalb2015} produce on-demand single photons, \cite{McKeever2004,Mucke2013} perform quantum gate operations, \cite{Reiserer2013,Reiserer2014} and to implement an elementary quantum network. \cite{Ritter2012} These works place single-atom quantum systems as a leading candidate for use in large-scale quantum networks.  

As a result, there is a strong interest in the integration of alkali atoms into robust, scalable, packaged optical cavities.\cite{Biedermann2010,Derntl2014}  Furthermore, it is desirable for these optical cavities to have small mode volumes and be tunable to atomic transitions.\cite{Trupke2007,Greuter2014,Flatten2015}  Here we report the development of `buckled-dome' Fabry-P\'{e}rot microcavities designed for cQED applications, specifically on-chip coupling between single photons and single rubidium atoms.  These cavities produce high single-atom cooperativities, can be easily tuned to atomic transitions, and can facilitate open-access for incorporation of atoms.

\begin{figure}[b]
	\includegraphics[width=8.5cm]{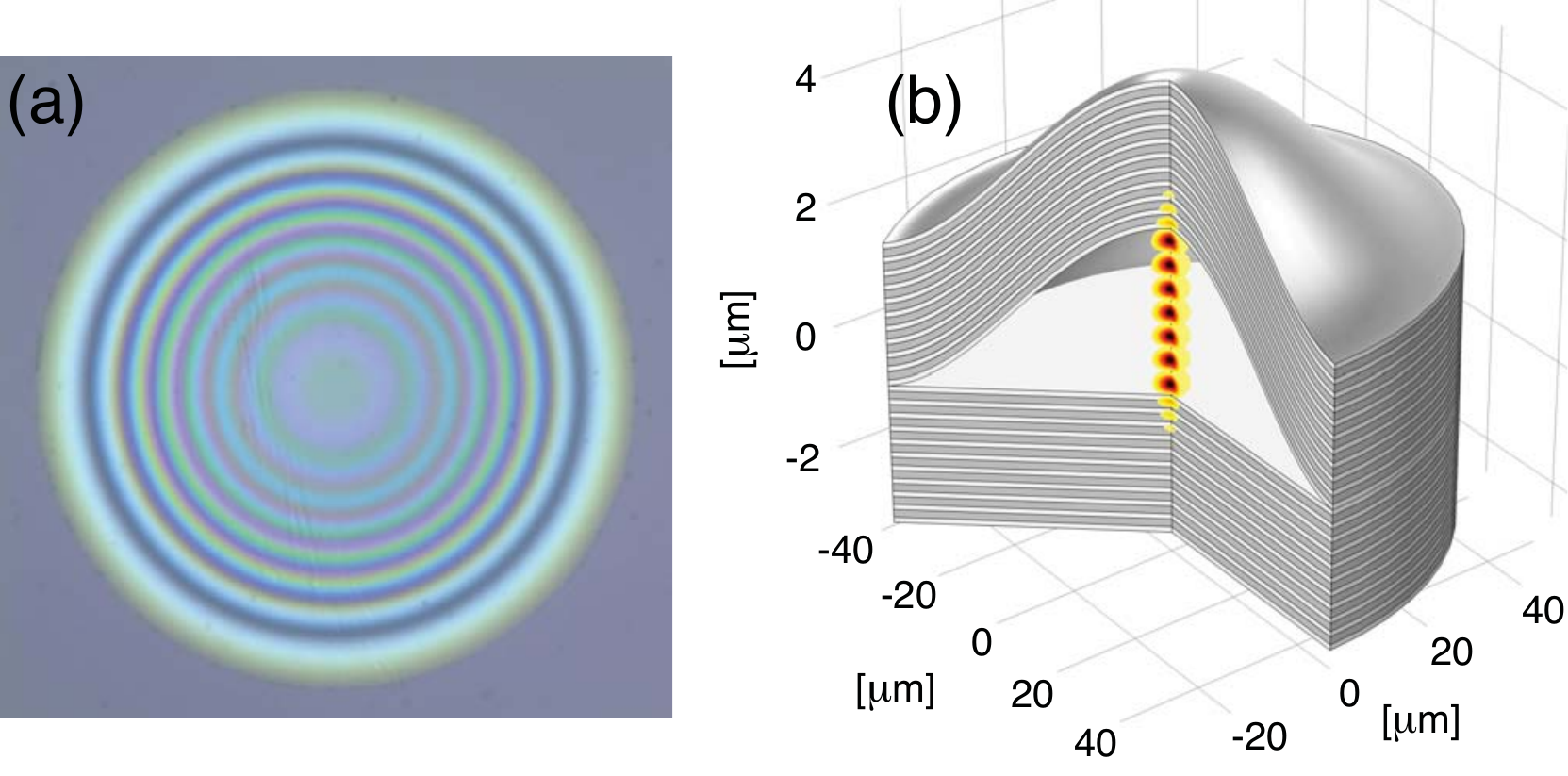}
	\caption{(a) Microscope image of a buckled-dome microcavity with base diameter of 100 $\mu$m. White-light interference rings demonstrate the high degree of cylindrical symmetry attained through the self-assembly process. (b) Cross-section of a finite element model of a 100 $\mu$m diameter microcavity (horizontal and vertical axes different scales) showing alternating SiO$_2$ (grey) and Ta$_2$O$_5$ (white) layers of the Bragg reflectors and a simulated optical mode.}
	\label{fig_1}
\end{figure}

The buckled-dome microcavities were fabricated via a monolithic self-assembly procedure. \cite{Epp2010,Allen2011} First, a distributed Bragg reflector (10.5 periods SiO$_2$/Ta$_2$O$_5$, starting and ending with Ta$_2$O$_5$) was deposited on a fused silica substrate by reactive magnetron sputtering. Microcavities were defined by the lithographic patterning of a thin ($\sim$15 nm) low-adhesion fluorocarbon layer, followed by the deposition of a second Bragg reflector identical to the initial reflector. Films with low loss and high compressive stress ($\sim$200 MPa) were realized by using high target power (200 W), elevated substrate temperature ($150\,^{\circ}$C), and low chamber pressure (4 mTorr). \cite{Hoffman1994} Optical constants for single films were measured using an ellipsometer. The refractive indices of SiO$_2$ and Ta$_2$O$_5$ were estimated to be 1.49 and 2.14 respectively, at a wavelength of 780 nm. At the same wavelength, extinction coefficients less than 10$^{-6}$ were estimated for both materials. The layer thicknesses were chosen such that all layers are nominally one quarter wavelength thick ($\lambda_0$/$4n$) at the D2 transition of $^{87}$Rb ($\lambda_0$ = 780.24 nm).\cite{Steck2003} Using a transfer matrix formalism,\cite{Hecht1987}  and the optical constants extracted from measurements on single films, a peak reflectance of $ 0.9991$ is predicted for the Bragg mirrors, corresponding to a reflection-limited cavity finesse of $\mathcal{F}$ = 3500.

Following the deposition of the top mirror, the samples were heated (400$\,^\circ$C) to induce a loss of adhesion between the two mirrors in the region of the fluorocarbon layer. The built-in compressive stress drives the release of the top mirror, forming a dome-like buckle. The height and exact morphology is dependent on the dome diameter.\cite{Allen2011,Bitarafan2015}  In this work microcavities with base diameters of 100 $\mu$m to 300 $\mu$m, Figure 1, and peak heights of $\delta = 2.5$ $\mu$m to 10 $\mu$m were studied. We describe in detail the 100 $\mu$m diameter cavities, and show select results for larger devices.  

\begin{figure}
	\includegraphics[width=8.5cm]{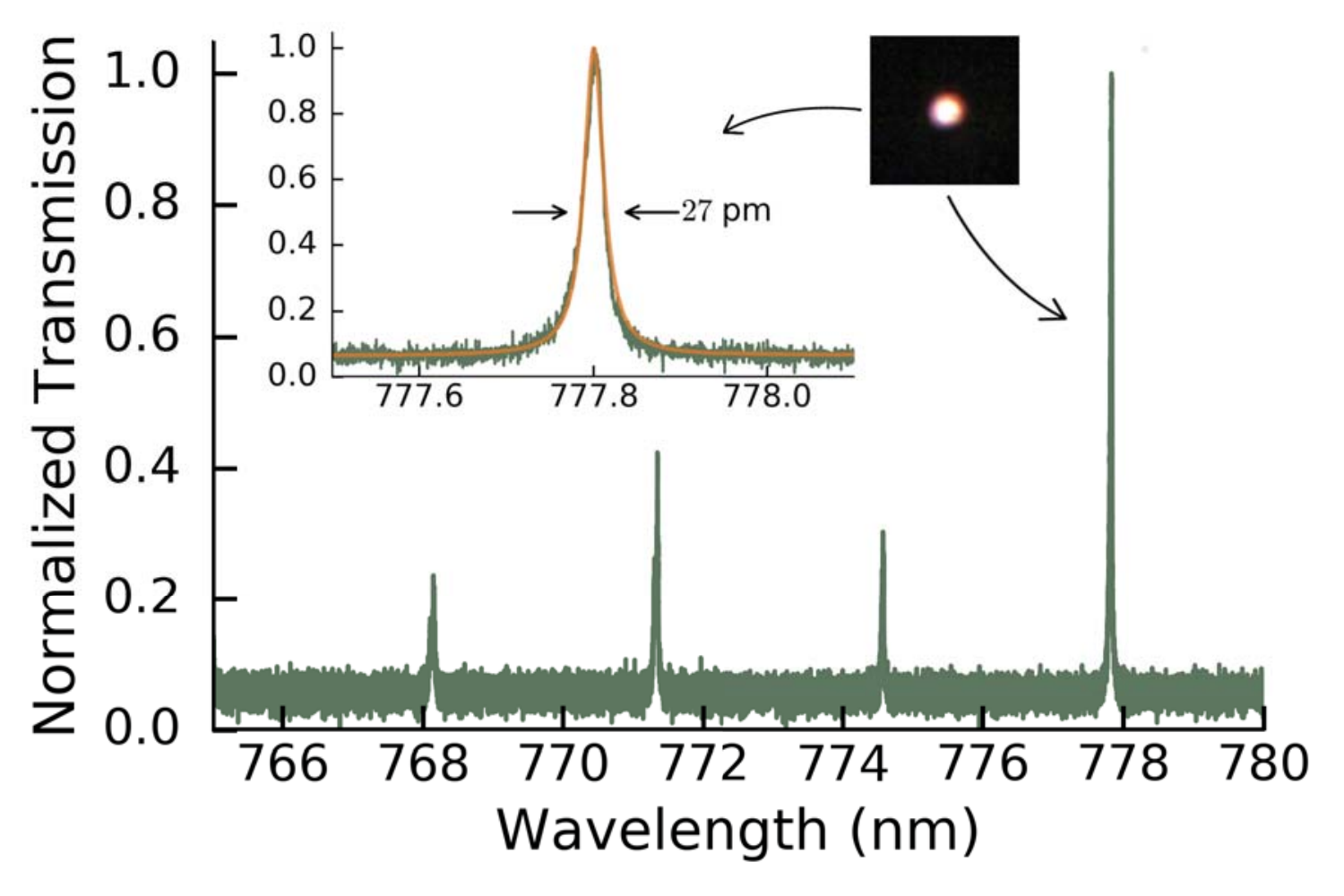}
	\caption{Transmission spectrum for a 100 $\mu$m diameter buckled-dome microcavity, with an optical cavity height of 2.67 $\mu$m.  The inset shows the fundamental mode resonance in greater detail, with a Lorentzian fit showing a full-width at half-maximum of 27 pm, corresponding to $\kappa = 2\pi \times 6.7$ GHz, and an image of the fundamental mode.}
	\label{fig_2}
\end{figure}

Optical resonances were examined by performing transmission spectroscopy using a fiber-coupled tunable diode laser (New Focus Velocity TLB6712), focused onto the microcavity using an objective lens ($50\times$ Mitutoyo Plan APO). Transmitted light was captured using a second objective lens ($100\times$ Mitutoyo Plan APO SL) and focused onto a photodiode (ThorLabs DET36A).  A typical wavelength sweep is shown in Figure 2, revealing resonances associated with the fundamental TEM$_{00}$ mode (at 777.80 nm) and three higher-order transverse modes. A digital camera was used to verify the profiles of these Laguerre-Gaussian modes. The fundamental mode was fit to a Lorentzian, revealing linewidths of $\kappa = 2\pi \times 2.7$ GHz for 300 $\mu$m diameter cavities and $\kappa = 2\pi \times 6.7$ GHz for 100 $\mu$m diameter cavities. Using the relation \cite{Hunger2010} $\kappa = \pi c/2L\mathcal{F}$, an estimate of the finesse may be made, where the effective cavity length is $L = \delta + 2d_p$, the geometric cavity length is $\delta$, and the penetration depth into the mirrors \cite{Hood2001}  is $d_p \approx (\lambda_0/2)(n_H-n_L)^{-1}$. For the case shown in Figure 2, this yields a finesse of $\mathcal{F} = 3560$, in good agreement with the predicted value.

The volume of the fundamental Gaussian mode of a Fabry-P\'erot cavity can be approximated as \cite{Hunger2010} $V_m = (\pi/4)w_0^2L$,  where $w_0$ is the mode waist (radius).  In the paraxial approximation for a half-symmetric Fabry-P\'erot cavity, the mode waist can be approximated as \cite{Hunger2010}
\begin{equation}
w_0 \approx \sqrt{\frac{\lambda}{\pi}}(R_c \times L)^{1/4},
\label{eqn_2}
\end{equation}
\noindent
where $R_c$ is the radius of curvature of the upper mirror and $L \ll R_c$ is assumed. $R_c$ was determined by fitting a circular segment to the profile of the optical cavity, as measured by optical profilometry (ZYGO MetroPro). The 100 $\mu$m buckled-dome cavity has a peak height of $\delta$ = 2.67 $\mu$m, a total cavity length of $L$ = 3.26 $\mu$m, and a radius of curvature of $R_c$ = 210 $\pm$ 15  $\mu$m. Eq.~\ref{eqn_2} then produces a mode waist of $w_0 \approx$ 2.55 $\mu$m, implying that $V_m \approx$ 35$\lambda^3$. This matches well with COMSOL simulations that yield a mode volume of 35.7$\lambda^3$. Similar values have been reported for other visible-wavelength optical cavities. \cite{Vahala2003,Muller2010,Dolan2010,Greuter2014} 

\begin{figure}[b]
	\includegraphics[width=8.5cm]{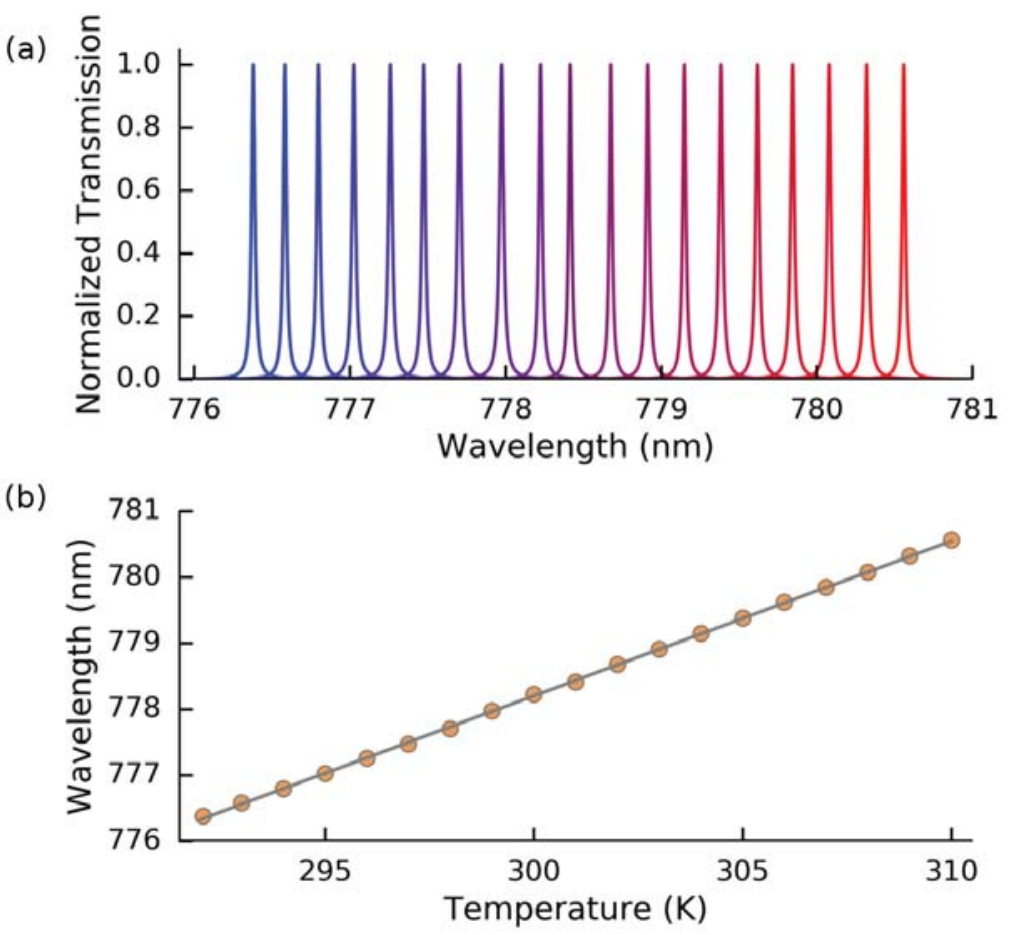}
	\caption{(a) Fits of the experimental resonance peaks for the fundamental mode of a 100 $\mu$m diameter dome, at temperatures from 292 K (blue, 776.4 nm) to 310 K (red, 780.6 nm) in 1 K increments. (b) The variation of the fundamental resonance wavelength with temperature. Orange circles are experimental data points, and the grey line is a linear fit yielding a thermal tunability of $\Delta \lambda/\Delta T$ = 0.2346 $\pm$ 0.0007 nm/K.}
	\label{fig_4}
\end{figure}

For a Fabry-P\'erot device to be considered viable for cQED applications, it should allow for practical tuning and stabilization of the resonance conditions. We have previously reported thermal tuning of the buckled-dome microcavities. \cite{Bitarafan2015b} Here samples were attached to a copper heat sink in a vacuum chamber at $\sim$5 mTorr, and the temperature was regulated by a proportional-integral-derivative controller. Transmission spectroscopy was performed in one degree intervals, as seen in Figure 3. The peak wavelength of the fundamental mode was tracked as a function of temperature, revealing a temperature dependence of $\Delta \lambda/\Delta T$ = 0.2346 $\pm$ 0.0007 nm/K. This tunability is vital because the stochastic nature of the buckling process produces cavities of varying resonance frequencies that must be drawn into resonance with the desired atomic transition. By heating the chip of microcavities, thermal expansion increases the cavity length sufficiently to align the optical resonance with the desired $^{87}$Rb transition at 308 K. The cavity wavelength could then, in principal, be locked to an atomic transition.  One drawback of the current method of tuning is the inability to tune individual cavities, however integrating heater electrodes\cite{Zalalutdinov2003} or electrostatic actuation\cite{Derntl2014} would allow for individually addressable on-chip microcavities.

Another requirement of a cQED device is that it should have open-access for injection of atoms into the cavity. The buckling self-assembly process inherently produces closed cavities, but can be modified or supplemented to provide open-access. One strategy is to couple the microcavities to hollow channels by patterning narrow strips of fluorocarbon that buckle along with the domes and intersect the microcavity.  A representative, $60$ $\mu$m wide, channel is shown in Figure 4a. This channel has a peak height of 2.7 $\mu$m, providing access for atomic gas injection. Previous work has demonstrated that mirror spacings as narrow as 110 nm were sufficient for atom injection,  \cite{Hood2001} and recent work has demonstrated vapor cells with critical dimensions as narrow as 30 nm. \cite{Whittaker2015} In addition, since these hollow channels are formed by two Bragg reflectors, they can also act as optical waveguides, possibly useful for atom trapping. 
\cite{Melnyk2015}  A second strategy for open access is to use focused ion beam (FIB) milling to remove small portions of the top Bragg reflector. Figure 4c shows an example of an access hole milled into a buckled-dome cavity.  

We found that the properties of the fundamental microcavity resonance were retained in the case of intersecting waveguides (see Figure 4b) likely because the optical mode-waist is small compared to the dome diameter. Thus, modifications of the dome periphery has minimal impact on the central part of the dome where the fundamental mode resides. However, we found that FIB milling resulted in microcavities with optical resonances that exhibit non-linear behavior (see Figure 4d), even at low input powers. This may be a result of the conductive carbon layer applied in order to avoid charging, or parasitic gallium implantation, during the FIB milling process.

\begin{figure}[t]
	\includegraphics[width=7.5cm]{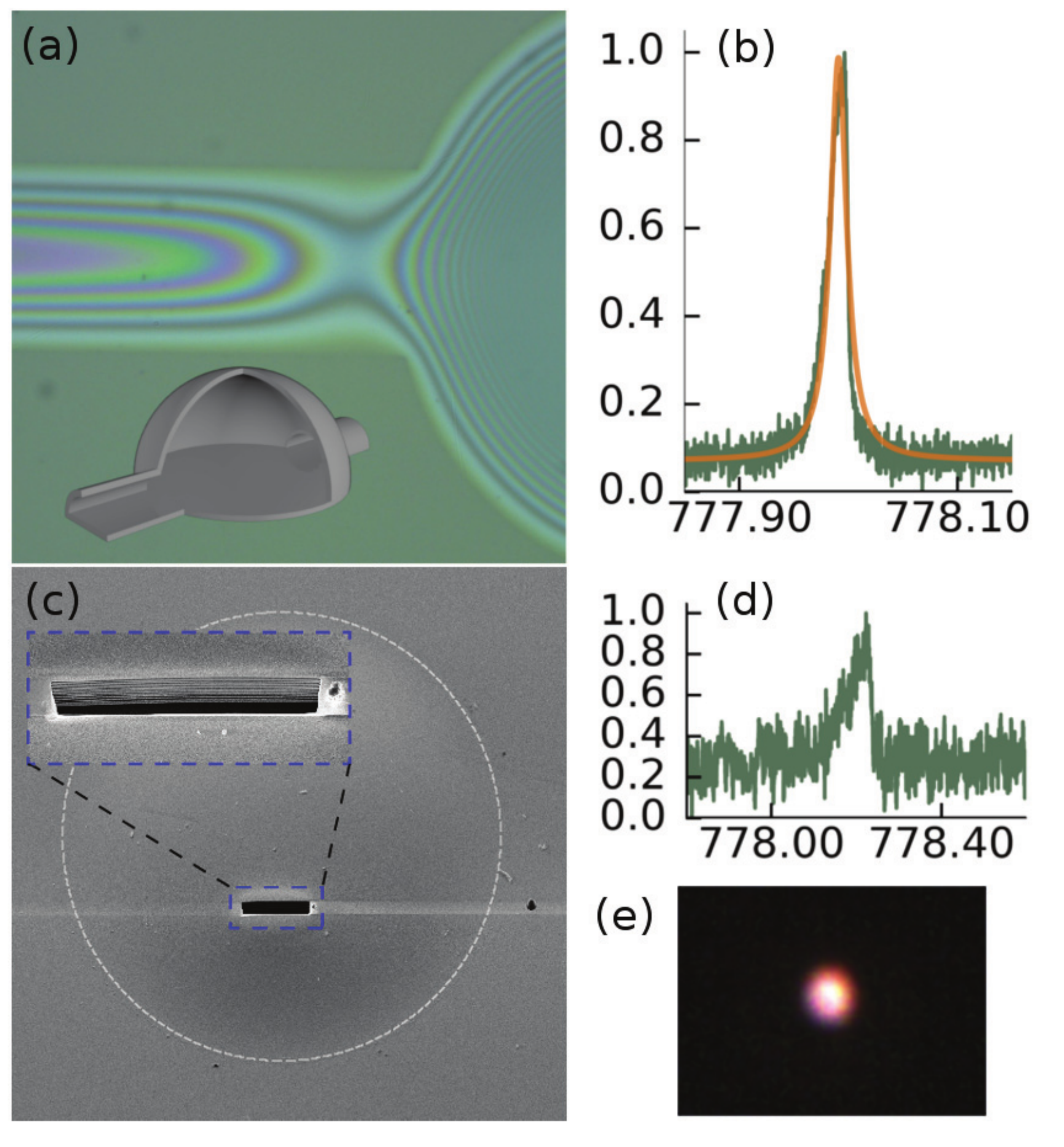}
	\caption{(a) Microscope image of a buckled-dome microcavity with a base diameter of 200 $\mu$m intersected by a 60 $\mu$m hollow channel. Inset is a 3D cartoon illustrating how the channel intersects the dome, to provide open access to the interior of the optical cavity. (b) Transmission spectra of the fundamental optical mode of dome cavity shown in (a). (c) SEM image of a 30 $\mu$m x 5 $\mu$m hole cut through the top mirror of a dome cavity via FIB milling. The open cavity is visible through the hole, as well as the distinct layers of the Bragg reflector. The cavity is outlined to improve visibility. (d) Transmission spectra and (e) image of the fundamental mode of the cavity shown in (c).  }
	\label{fig_5}
\end{figure}

Finally we consider some pertinent parameters for cQED applications: the atom-cavity coupling rate $g_0$, and the cavity decay rate $\kappa$. For a cavity with resonance frequency $\omega$, and a single atom located at the maxima of the cavity field, $g_0$ is defined as
\begin{equation}
g_0 = \sqrt{\frac{3\lambda^2c\gamma}{4\pi V_m}},
\label{eqn_1}
\end{equation}
\noindent
where $\gamma$ is the half-width at half-maximum linewidth of the excited state of the atom ($\gamma = 2\pi \times 3.0$ MHz for $^{87}$Rb), and $V_m$ is the optical mode volume. To maximize $g_0$, the mode volume must be minimized. A mode volume of 35$\lambda^3$ results in a coherent atom-cavity coupling rate of $g_0 = 2\pi \times 1.12$ GHz, greater than the most optimistic predictions for macroscopic Fabry-P\'erot cavities, \cite{Hood2001} and among the highest reported to date in the literature. \cite{Hunger2010}

\begin{figure}[t]
	\includegraphics[width=8.0cm]{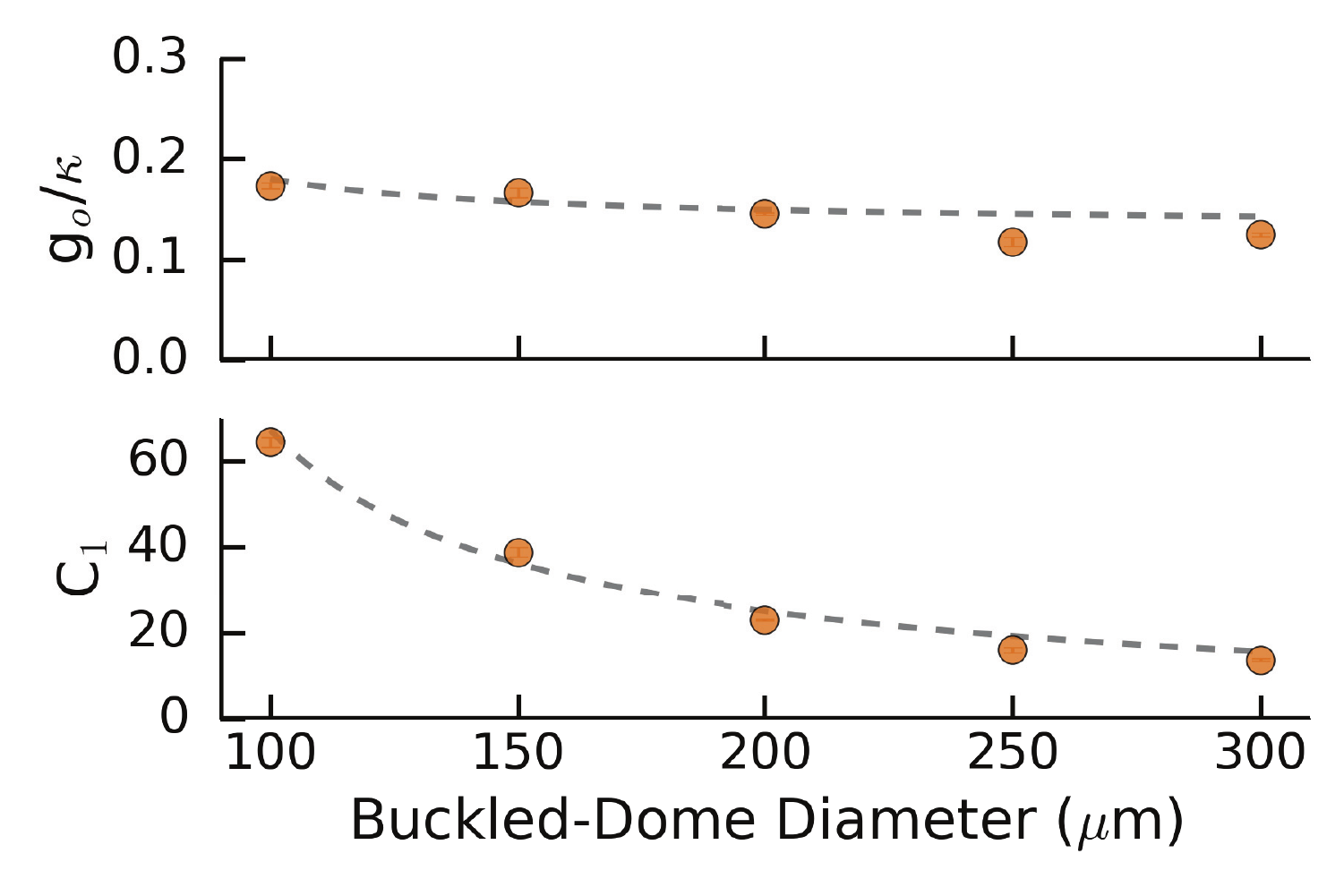}
	\caption{(a) Strong coupling parameter, $g_0/\kappa$, as a function of dome diameter. Fit (grey dashed line) derived using cavity finesse and optical profilometry measurements of dome height and radius of curvature. Deviations from the predicted zero-slope behavior are a result of the divergence from the elastic buckling model, producing a non-linear dependence between the radius of curvature and the dome diameter. \cite{Bitarafan2015} (b) Experimental cooperativity as a function of dome diameter.  Fit as described in (a). Error bars, from the statistical uncertainty in $\kappa$ and measurement uncertainty in $R_c$, are smaller than the symbol size.}
	\label{fig_3}
\end{figure}

Despite the high atom-cavity coupling rate, a more relevant figure of merit for cQED applications is the strong coupling parameter $g_0/\kappa$. \cite{Hood1998} The cavities described here exhibit $g_0/\kappa \leq 0.17$ placing them in the weakly coupled regime of cQED, which as outlined below could reasonably be improved.  Nonetheless, it is informative to consider the dependence of this ratio on the diameter of the buckled-dome cavity. From above it can be seen that
\begin{equation}
\frac{g_0}{\kappa} \propto   \frac{L}{\sqrt{V_m}},
\label{eqn_3}
\end{equation}
\noindent
where $V_m \propto R_c^{1/2}L^{3/2}$. Furthermore, it has been shown previously,\cite{Allen2011} that based on elastic buckling mechanics $L \propto D$, and to first order $R_c \propto D$ (where $D$ is the buckled-dome diameter). Thus, Eq.~\ref{eqn_3} predicts  ${g_0}/{\kappa} \propto {D}/{\sqrt{D^2}}$.  That is, to first order the strong coupling parameter is independent of the buckled-dome diameter.  Experimentally we observe a slight increase in the strong coupling parameter with \textit{decreasing} dome diameter, as shown in Figure 5a. This behavior is in contrast to the typical trend of an improving strong coupling parameter with \textit{increasing} cavity length. \cite{Hunger2010} 

The ability to maintain a constant strong coupling parameter while decreasing the mode volume is important for another key figure of merit of cQED, $C_1$, known as the single-atom cooperativity, which we define following Law \textit{et al.}\cite{Law1997} as
\begin{equation}
C_1 = \frac{g_0^2}{\kappa\gamma}.
\label{eqn_4}
\end{equation}
\noindent
$C_1$ is of particular interest for single-photon sources as it defines the Purcell factor; the probability of a spontaneously emitted photon entering the cavity mode is given by $2C_1/(2C_1+1)$. It should be noted that both $\kappa$ and $\gamma$ are defined as the half-width at half-maximum. Given the parameters of the buckled-dome cavities, Eq.~\ref{eqn_4} produces $C_1 = 65$ (Figure 5b).  For comparison, cooperativities as high as $C_1 = 290$ and $51$ have been reported for fiber-based cavities\cite{Colombe2007} and macroscopic cavities, \cite{Sauer2004} respectively.  

In conclusion, we have developed on-chip Fabry-P\'erot microcavities specifically for cQED applications using alkali atoms, with tunable resonance frequencies and open-access for atom inclusion. Despite only moderate atom-photon coupling ($g_0/\kappa = 0.17$), high single-atom cooperativity ($C_1 = 65$) provides strong motivation for further investigation and optimization. The results presented in this letter were obtained with ten-period Bragg reflectors, producing reflection limited finesse cavities with $\mathcal{F} = 3560$. Increasing the number of periods would increase reflectance, and macroscopic cavities using the same materials have reported finesses as high as $\mathcal{F}$ = 480,000. \cite{Hood2001} In order to achieve strong coupling ($g_0/\kappa \geq 1$) with our buckled-dome Fabry-P\'erot cavities, a finesse of $\mathcal{F} \geq$ 21,000 would be required. This is modest compared to finesse requirements predicted for other optical cavity architectures. \cite{Ritter2012,Hunger2010,Mabuchi1999} Furthermore, improvement of the finesse to this level would have the benefit of increasing $C_1$ as high as 420, making it among the highest cooperativity architectures available.  

We are grateful for the support of Alberta Innovates Technology Futures (Strategic Chair \& iCiNano); the University of Alberta Faculty of Science; Natural Sciences and Engineering Research Council, Canada (RGPIN 401918 \& EQPEG 458523); the Canada Foundation for Innovation; and the Alfred P. Sloan Foundation. LJL acknowledges that this research was undertaken, in part, thanks to funding from the Canada Research Chairs program.

\end{document}